\documentclass[12pt,preprint]{aastex}
\usepackage{graphicx}
\usepackage{emulateapj5}
\usepackage{apjfonts}
\usepackage{epsfig}
\usepackage{natbib}
\newcommand{\err}[2]{\ensuremath{^{_{+#1}}_{^{-#2}}}}
\newcommand{\ee}[2]{\ensuremath{{#1}\!\times\!10^{#2}}}

\newcommand{\ergcms}{\ensuremath{\mathrm{erg~cm}^{-2}~\mathrm{s}^{-1}}}
\newcommand{\pcmsq}{\ensuremath{\mathrm{cm}^{-2}}}

\newcommand{\chandra}{\emph{Chandra}}

\def\gtrsim{\mathrel{\hbox{\rlap{\hbox{\lower4pt\hbox{$\sim$}}}\hbox{\raise2pt\hbox{$>$}}}}}

\newcommand{\hst}{\emph{HST}}

\newcommand{\msun}{\ensuremath{M_{\odot}}}

\newcommand{\oiii}{[\ion{O}{3}]}

\newcommand{\bub}{SDSS~J1356$+$1026}

\def\lax{{$\mathrel{\hbox{\rlap{\hbox{\lower4pt\hbox{$\sim$}}}\hbox{$<$}}}$}}
\def\gax{{$\mathrel{\hbox{\rlap{\hbox{\lower4pt\hbox{$\sim$}}}\hbox{$>$}}}$}}

\slugcomment{April 18, 2014; accepted by {\it The Astrophysical Journal}.}
\shorttitle{{\it X-ray Outflow}}
\shortauthors{GREENE, ET AL.}

\begin{document}

\title{Extended X-ray Emission From a Quasar-Driven Superbubble}

\author{Jenny E. Greene\altaffilmark{1}, 
David Pooley\altaffilmark{2}, Nadia L. Zakamska\altaffilmark{3}, 
Julia M. Comerford\altaffilmark{4}, Ai-Lei Sun\altaffilmark{1}}

\altaffiltext{1}{Department of Astrophysics, Princeton University, Princeton, NJ 08540}
\altaffiltext{2}{Department of Physics, Sam Houston State University, Huntsville, TX  77341,
Eureka Scientific, Inc., 2452 Delmer Street Suite 100, Oakland, CA  94602}
\altaffiltext{3}{Center for Astrophysical Sciences, Department of Physics and 
Astronomy, Johns Hopkins University, Baltimore, MD 21218, USA}
\altaffiltext{4}{Department of Astrophysical and Planetary Sciences, University of Colorado, Boulder, CO 80309, USA}

\begin{abstract}
  We present observations of extended, 20-kpc scale soft X-ray gas
  around a luminous obscured quasar hosted by an ultra-luminous
  infrared galaxy caught in the midst of a major merger.  The extended X-ray emission
  is well fit as a thermal gas with a temperature of \emph{kT}$\approx 280$
  eV and a luminosity of $L_{\rm X}\approx 10^{42}$ erg~s$^{-1}$ and
  is spatially coincident with a known ionized gas outflow.  Based on
  the X-ray luminosity, a factor of $\sim 10$ fainter than the \oiii\
  emission, we conclude that the X-ray emission is either
  dominated by photoionization, or by shocked emission from cloud surfaces in
  a hot quasar-driven wind.
\end{abstract}

\section{Introduction}

Black hole (BH) ``feedback'' is often invoked to regulate massive
galaxy formation \citep[e.g.,][]{springeletal2005}, but definitive
examples of radiation from the quasar accretion disk driving powerful
outflows are difficult to find \citep[e.g.,][]{moeetal2009}.  While clear
evidence exists that powerful radio jets entrain warm gas and carry
significant amounts of material out of their host galaxies
\citep[e.g.,][]{nesvadbaetal2006,fustockton2009}, the situation is far
less clear for radio-quiet targets, which dominate the active galactic
nucleus (AGN) population. In the past few years, we have identified a
number of radio-quiet quasars with high intrinsic luminosities ($M_B <
-26.9$ mag) that show outflowing ionized gas on $\sim 15$ kiloparsec (kpc) scales
\citep{greeneetal2011, greeneetal2012, liuetal2013a,liuetal2013b,
  hainlineetal2013}. In this paper, we present observations of hot
X-ray gas that is aligned with a known 20-kpc--scale ionized gas
bubble in the radio-quiet quasar SDSS J135646.10+102609.0 
(SDSS J1356+1026 hereafter).

SDSS\,J1356$+$1026 is an on-going merger and ultra-luminous infrared
galaxy (ULIRG) located at $z$=0.123 ($D_L = 568$ Mpc). The Northern
nucleus hosts a luminous ($L_{bol}\simeq 10^{46}$ erg s$^{-1}$)
obscured quasar that was originally discovered in the Sloan Digital
Sky Survey \citep[SDSS;][]{yorketal2000} based on its 
\oiii$~\lambda 5007$ emission \citep{zakamskaetal2003,reyesetal2008}.
The source was also flagged as a possible dual active nucleus, because of the
presence of multiple velocity components in the SDSS spectrum,
although that classification has been questioned
\citep{liuetal2010,fuetal2012}.

Our interest here lies in the $\sim$20 kpc outflow that we discovered in
long-slit observations with Magellan \citep[][Paper I
hereafter]{greeneetal2009,greeneetal2012}.  In our Magellan spectrum,
we detect the line splitting that is characteristic of an expanding
``bubble'' (Figure \ref{fig:xrayimage}). The shell region extends $\sim 10$
kpc ($\sim 4$\arcsec) to the South of the quasar hosted by the Northern nucleus.
Equidistant from the Northern nucleus to the North are clumps of
\oiii\ emission with comparable observed velocities to the bubble.  We thus
propose that we are observing an expanding bipolar super-bubble
similar to those observed in many star-forming galaxies both with and
without AGNs \citep[e.g.,][]{heckmanetal1990,rupkeveilleux2013}.

\section{A Quasar-Driven Superwind?}

As discussed in Paper I, based on the velocity splitting in our
long-slit spectrum, we find that the bubble is expanding symmetrically
about the Northern nucleus with deprojected velocities of $\sim 1000$
km s$^{-1}$. We infer that the outflow is driving a shell of ionized
gas with a kinetic energy of ($\sim$10$^{44}$--10$^{45}$ erg
s$^{-1}$), and we argue that accretion energy is the most likely
energy source driving the outflow.  We consider alternative sources to
power the outflow, such as a relativistic jet or star formation in
the host galaxy, and conclude that neither of these alternatives is
energetic enough to produce the observed structures.  Since \bub\ is
radio-quiet, with a radio source that is unresolved by FIRST
\citep{beckeretal1995}, we think it unlikely that a kpc-scale jet is
driving the emission (although deeper radio data should be
illuminating, Wrobel et al. in prep).  

Converting the observed far-infrared (FIR) luminosity into a star
formation rate of $\sim 100$~\msun~yr$^{-1}$ yields a kinetic
luminosity of $\sim 10^{44}$ erg~s$^{-1}$
\citep{leithererheckman1995,veilleuxetal2005}, at the low end of our
estimates.  However, the star formation rate is likely far lower than
this.  Based on the gas surface density inferred from ALMA
observations of CO in \bub, combined with a fit to the far-infrared 
spectral energy distribution, we find a star formation rate that is
$<21$~\msun~yr$^{-1}$ under a number of conservative assumptions 
(A. Sun et al. submitted).  Furthermore, the high level of $8\%$ optical
polarization that we present in Paper I argues strongly that the AGN
dominates the blue light at optical wavelengths.

\begin{figure*}
\vbox{ 
\vskip -5mm
\hskip +13mm
\includegraphics[width=0.85\textwidth]{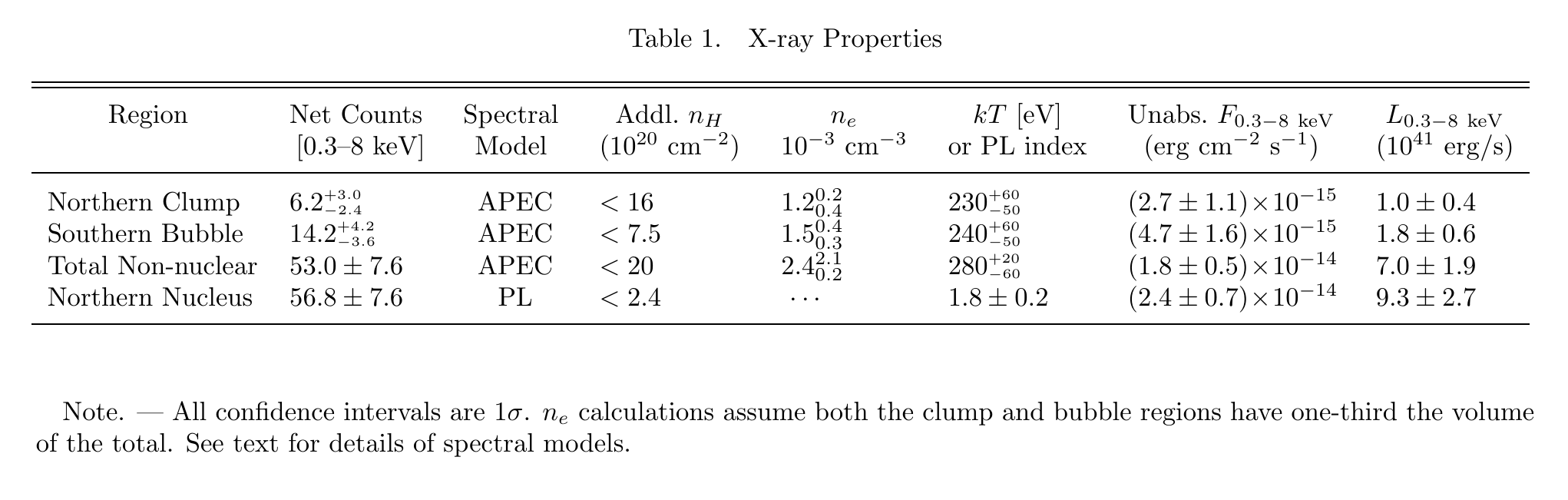}
}
\vskip -0mm
\end{figure*}

\begin{figure*}
\includegraphics{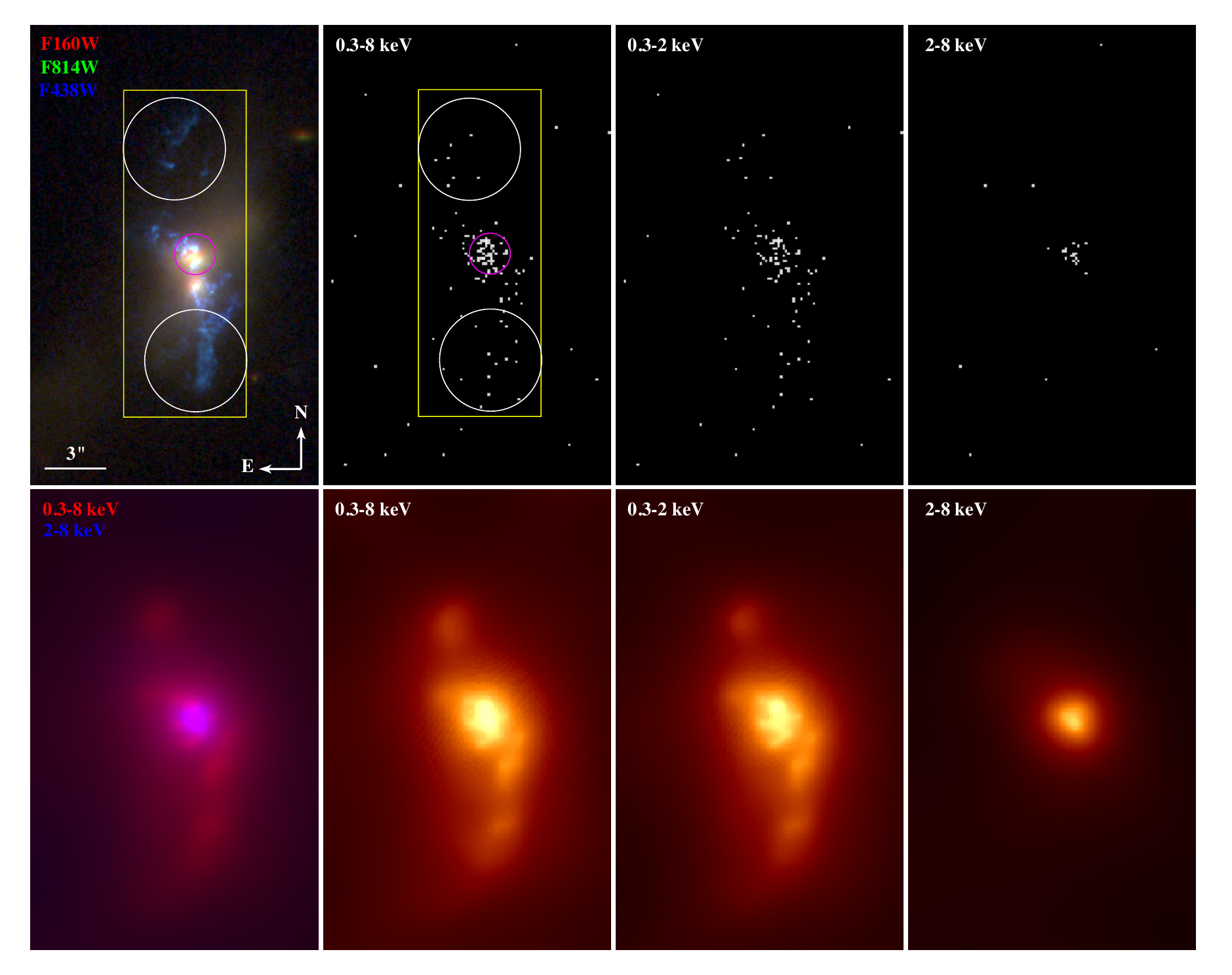}
\caption{\hst\ (top left) and \chandra\ views of \bub.  In the top-left panel, 
we show the three-band WFC3 composite image.  The extended blue/green
light is dominated by line emission, while the red is stellar tidal features. 
The orientation of all images has North up and East to the left, and the extraction 
regions for the bubble and clumps are indicated by white circles, while the total 
extraction region is shown as a yellow rectangle and the North nucleus extraction 
region is the magenta circle. 
In the remaining top panels, we show the X-ray maps in the 0.3--8, 0.3--2, 
and 2--8 keV energy ranges. In the bottom panels, we show adaptively smoothed 
images to guide the eye, although we caution against overinterpreting these.
\label{fig:xrayimage}
}
\end{figure*}

Our preferred model, as described in Paper I, is that radiative energy
from the AGN itself is driving a shocked wind into the interstellar
medium of the galaxy, and further out into the intergalactic
medium. The ionized gas we observe is just the frosting on a
predominantly hot outflow
\citep[e.g.,][]{fauchergiguerequataert2012,zubovasking2012}.  In
starburst galaxies, overlapping supernova remnants combine to create a
hot expanding bubble of gas that, if sufficiently overpressured, will
``break out'' of the galaxy along the minor axis
\citep[e.g.,][]{chevalierclegg1985,maclowmccray1988}.  These hot winds
can power kpc-scale superbubbles that are observable in the soft X-ray
\citep[e.g.,][]{fabbianoetal1990,stricklandetal2004a,stricklandetal2004b},
as well as ionized \citep[e.g.,][]{heckmanetal1990,rupkeveilleux2013},
and neutral \citep[e.g.,][]{rupkeetal2005} gas \citep[see][for a
recent review]{veilleuxetal2005}.  While we believe AGNs also drive
hot winds \citep[e.g.,][]{choietal2013}, they are much more
challenging to unambiguously detect in the X-rays, due to confusion
from photoionization \citep[e.g.,][]{bianchietal2006,wangetal2011},
star formation \citep[e.g.,][]{wangetal2009}, and collisional
ionization \citep[e.g.,][]{paggietal2012}.  \bub\ provides an exciting
opportunity to explore the nature of X-rays from an AGN-driven
superbubble on $\sim 20$ kpc scales.

\section{X-ray Observations and Analysis}

J. Comerford et al.\ in prep.\ observed \bub\ for 19.8 ks on 2012
March 31 (ObsID 13951) with the Advanced CCD Imaging Spectrometer
(ACIS). The data were taken in timed-exposure mode with an integration
time of 3.14 s per frame, and the telescope aim point was on the
back-side illuminated S3 chip. The data were telemetered to the ground
in faint mode.

Reduction was performed using the CIAO 4.5 software
\citep{fruscioneetal2006} provided by the \chandra\ X-ray
Center. The data were reprocessed with the chandra\_repro script using
the CALDB 4.5.8 set of calibration files (gain maps, quantum
efficiency, quantum efficiency uniformity, effective area).  The data
were filtered using standard event grades and excluding both bad
pixels and software-flagged cosmic-ray events. No intervals of strong
background flaring were found.

We use 5$''$ diameter regions (see Fig.~\ref{fig:xrayimage}) to
extract counts from the northern clump region and the southern bubble
region, and we use a 2$'$ source-free region to the southwest to
estimate the background contribution, which is small (less than one
count for the clump and bubble regions).  Finally, as an upper limit
to the extended diffuse X-ray counts, we extract all counts within a
$6 \times 16\arcsec$ region centered on the Northern nucleus,
excluding a 1\arcsec\ region around the Northern nucleus itself.  We
present the extracted counts in each region in Table 1.  Because there
are only a small number of counts in each source region, we use the
Bayesian method of \citet{kraftetal1991} to estimate the uncertainties
on the net counts.

Spectra were extracted from the clump, bubble, Northern nucleus, and
total non-nuclear regions, along with a background spectrum from the
2$'$ background region.  The unbinned spectra were fit in Sherpa
\citep{freemanetal2001} using \citet{cash1979} statistics.  The
spectral model for the clump, bubble, and total extended regions was
an absorbed \citep{wilmsetal2000} thermal plasma model (APEC) at the
redshift of the host galaxy.  The confidence intervals for the
parameters are determined using the task conf within Sherpa, taking
into account the presence of several variables in the fit.  

\begin{figure*}
\vbox{ 
\vskip -9mm
\hskip +3mm
\includegraphics{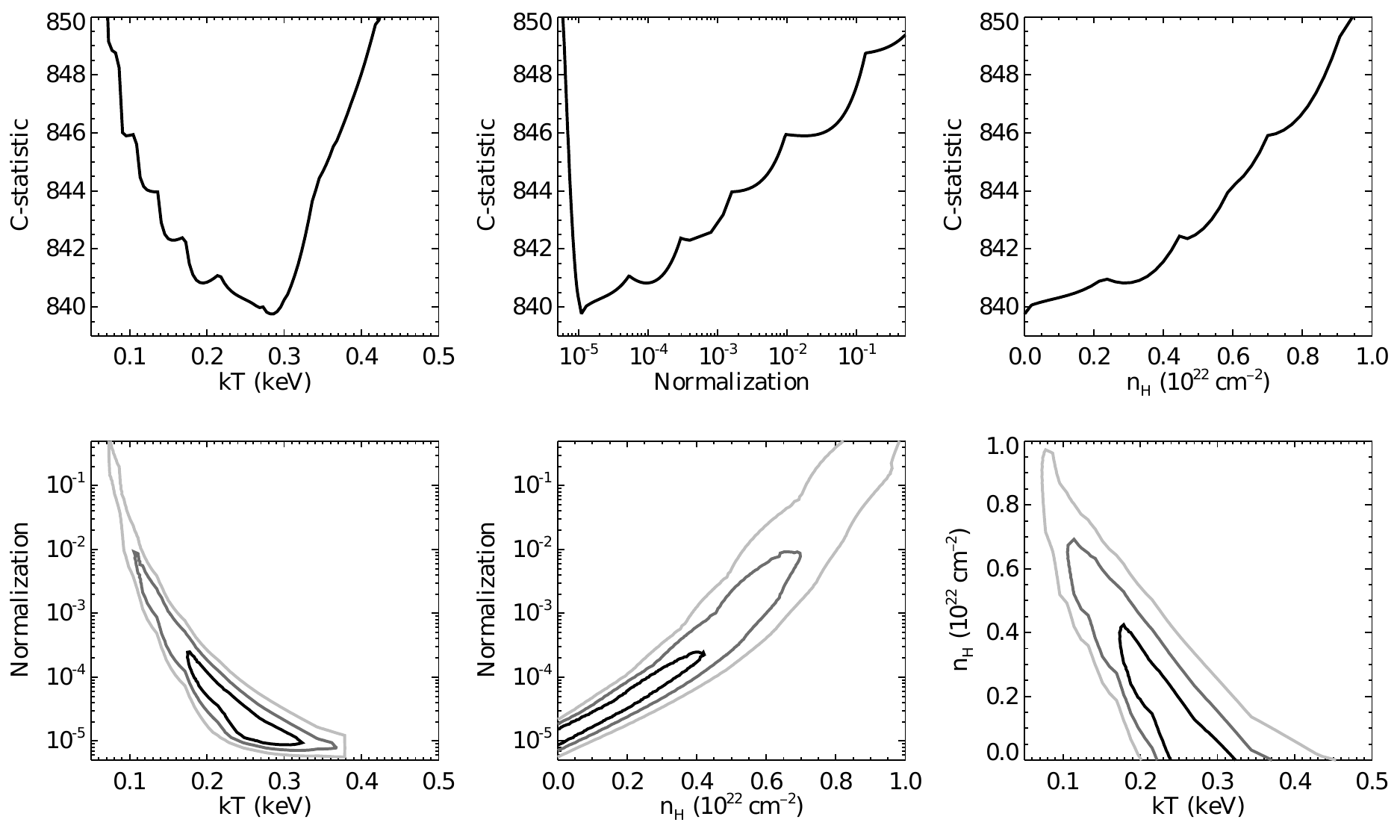}
}
\caption{{\it Top}: 
Cash statistic for each parameter of interest derived by fixing the latter 
and allowing all other parameters to vary freely.  The normalization 
is defined to be $(10^{-14} / 4 \pi [D_A (1 + z)]^2)
\int{n_e n_H dV}$ where $D_A$ is the angular diameter distance to the source 
in cm and $n_e$ and $n_H$ are the electron and hydrogen 
density in cm$^{-3}$. There is a distinct minimum in 
both $kT$ and normalization.  {\it Bottom}: Two-dimensional confidence contours 
derived in a similar fashion, by fixing the two parameters of interest over a grid and then 
allowing the fit to converge at the best Cash statistic.  The contour levels connect the 
$1~\sigma$, $2~\sigma$, and $3~\sigma$ regions where the Cash statistic has a value 
2.30, 6.18, and 11.83 higher than the overall best model.
\label{fig:showfit}
}
\end{figure*}

We fix the abundances to their solar values for our fiducial fit.  In
Figure \ref{fig:showfit}, we show the Cash statistic minimum for each
fitted parameter: $kT$, normalization, which scales with gas density,
and absorbing column $n_H$. Each parameter of interest $p$ is stepped
through a grid of 100 points between the limits shown, and $p$ is held
fixed while all other parameters are allowed to vary to find the best
fit at this value of $p$.  The Cash statistic of this new best fit is
plotted at each value, showing that we find clearly defined minima in
both the temperature $kT$ and the normalization.  The two-dimensional
confidence contours are created in a similar way, and plotted at the
$1, 2, 3~\sigma$ levels.  Within the context of a thermal plasma
model, $kT$ is very low no matter the value of the other parameters.
This low temperature can be seen from the X-ray counts themselves; the
majority of the extended counts are below 1 keV. It may be that the
assumption of solar abundances is incorrect; \citet{nardinietal2013}
find abundances closer to half the solar value in the X-ray halo of
NGC 6240, perhaps with a component of one-tenth solar metallicity as
well. As an additional check, we also fit models with abundances
one-half and one-tenth solar and find differences in the best-fit
\emph{kT} only at the $<10\%$ level.

The absorbing column consists of two components,
one fixed at the Milky Way value of $n_H = \ee{1.9}{20}~\pcmsq$ in the
direction of SDSS J1356+1026 \citep{dickeylockman1990}, and an
additional component at the redshift of SDSS J1356+1026.  No
additional absorption was required, with $1 \, \sigma$ upper limits
given in Table 1. We calculate the unabsorbed fluxes from the best-fit
models.  To estimate the uncertainties in the flux, we used the
sample\_energy\_flux command in Sherpa to obtain 10\,000 samples of
the energy flux, taking into account the uncertainties in all of
the spectral parameters. The uncertainties are dominated by statistical 
uncertainties at these flux levels, and we use the standard deviations of these
sets of samples as the uncertainties for the fluxes (Table 1).  
While we are fitting a low total number of counts, the fact that the 
photons are all at very low energies rules out both high temperatures 
and large absorbing columns. The agreement between all three region 
fits provides extra support for our methodology and derived 
uncertainties. 

The North nucleus itself was fit with an absorbed power-law model
[with slope $\Gamma$, $N(E) \propto E^{\Gamma}$] with the same
absorption components as above.  Surprisingly, we find no evidence for
intrinsic absorption in the X-ray spectrum, although the source is
classified as a narrow-line AGN in the optical and we have detected
8\% polarization, usually a tell-tale sign of an obscured AGN (Paper I).
The best-fit power-law slope is $1.8 \pm 0.2$, and the column-density
settled at the minimum value, with a $1 \, \sigma$ upper uncertainty
of $1.8 \times 10^{20}$~cm$^{-2}$. This is a standard power-law slope
for local unobscured Seyfert galaxies
\citep[e.g.,][]{jinetal2012}. Perhaps the source is similar to
``changing-look'' AGNs, which transition from Compton-thick to thin,
but are obscured in the optical \citep[e.g.,][]{risalitietal2005}.

\subsection{Optical Line Emission}

In determining the nature of the X-rays, we compare their morphology
and luminosity of the warm ionized gas, as traced by \oiii\ in our
long-slit Magellan data \citep{greeneetal2009} and three-band
\hst/WFC3 imaging in F435W ($B-$band), F814W ($I-$band), and F160W
($H-$band) from Comerford et al.\ in prep. The total \oiii\ luminosity
within our slit is $\sim 10^{43}$~erg~s$^{-1}$. Because the line
emission is spatially extended, this estimate is strictly a lower
limit. Given that the majority of the emission visible in the \hst\
images align with our slit, we expect we are within a factor of two to
three of the total \oiii\ emission.  From our long-slit data, we
determine that the nuclear emission within 1\arcsec\ of the North
nucleus accounts for only $\sim 10\%$ of the total flux, a small
fraction compared to the larger (and more uncertain) aperture
corrections.  Thus, we consider the extended \oiii\ emission to have a
luminosity of $L_{\rm [OIII]}\approx 10^{43}$~erg~s$^{-1}$.  We trace
the overall line morphology using the \hst/WFC3 F438W and F814W
images, because the line emission has high enough equivalent width to
be observed in a broad-band image (Figure 1).  The morphology of the
X-rays matches that of the line emission, to the extent we can tell
from these relatively shallow observations.

\section{Discussion}

We find a total soft X-ray luminosity of $5-9\times
10^{41}$~erg~s$^{-1}$ with a best-fit temperature of $\sim 280$ eV,
and a total \oiii\ luminosity in the extended component of $\sim
10^{43}$~erg~s$^{-1}$.  Thus the extended X-ray emission is roughly an
order of magnitude fainter than the extended \oiii\ emission.  We
discuss below the possible origins of the overwhelmingly soft extended X-ray
emission. The prime suspects, in order from least to most probable,
are electron scattering, superwind-driven shocks, and photoionization.
We note that \citet{mcdowelletal2003} also suggest merger-driven
shocks as the heating mechanism of the X-ray emitting gas in Arp 220 
\citep[see also][ for similar arguments about NGC 6240]{nardinietal2013}.
However, as pointed out by \citet{grimesetal2005}, non-starbursting
mergers do not appear to show similar X-ray halos, strongly suggesting
winds over mergers as the heating mechanism.

\subsection{Electron scattering}

Electron scattering of the AGN continuum could provide X-ray photons,
but cannot account for the observations.  For one thing, scattering
would not change the hard spectral slope intrinsic to the AGN, so
cannot explain the very soft extended X-rays.  Secondly, the
scattering efficiency is too low to account for the observed
luminosity.  If we assume that the intrinsic UV luminosity is $\sim$
twice that of the infrared bump in Paper I
\citep[e.g.,][]{richardsetal2006}, we can estimate an intrinsic $\nu
L_{\nu}$[2000~\AA$]=3\times 10^{45}$ erg~s$^{-1}$.  Assuming that all
UV emission is due to electron-scattered light from the nucleus places
an upper limit on the electron-scattering efficiency of 3\% (much less
so if there is a contribution from dust). Since electron-scattering is
wavelength independent and dust scattering is negligible at X-ray
wavelengths, and given the luminosity of the North nucleus $L_{\rm X}
= \ee{(9 \pm 3)}{41}$~erg~s$^{-1}$, an upper limit to the scattered
X-ray luminosity is $\ll 3\% \times (9 \pm 3) \times 10^{42}$
erg~s$^{-1}$, or $L_{\rm X}/L_{\rm [OIII]} \ll 0.003$, much lower than
the observed X-ray luminosity. Finally, the amount of scattering is
constrained by the amount of line emission.  From
\citet{zakamskaetal2005}, if we take $n_e \sim 100$~cm$^{-3}$ and $d
\sim 5$~kpc, we expect $L_{\rm X}/L_{\rm [OIII]} \approx 0.005$, also
lower than the ratio of $\sim 0.1$ that is observed.

\subsection{Superwind}

Perhaps the most exciting possibility is that we are detecting shocked
gas associated with a wide-angle, quasar-driven wind
\citep[e.g.,][]{fauchergiguerequataert2012,choietal2013}.  Such a
super-wind would have similar properties to those blown by starbursts,
but the gas would be heated by the accreting black hole rather than
star formation.

To estimate the expected luminosity in the soft X-rays from a superwind, 
we follow \citet{heckmanetal1996} and assume that the
expanding bubble of hot gas behaves like a supernova remnant, with
negligible radiative losses, in order to translate our estimated
mechanical luminosity of $L_{\rm mech} \approx
10^{44}-10^{45}$~erg~s$^{-1}$ into an expected X-ray luminosity
\citep[see also][]{chevalierclegg1985}.  Their derived relationship is:
\begin{equation}
L_{\rm X} \approx 
3.1 \times 10^{40} L_{\rm mech, 43}^{33/35} \, n_{e,-2}^{17/35} \, t_7^{19/35} \, {\rm erg\, s^{-1}},
\end{equation}
where $L_{\rm mech}$ is in units of $10^{43}$~erg~s$^{-1}$, 
$n_{e,-2}$ is the density in units of $10^{-2}$~cm$^{-3}$ and $t_7$ is
the estimated age of the bubble in units of $10^7$ yr.  The observed
X-ray luminosity has $L_{\rm X} \approx f \, n_e \, n_H V \epsilon$,
with APEC calculating the emissivity $\epsilon$, $f$ the volume filling factor,
$n_e \approx n_H$, and the volume $V$ assumed to be a cylinder with
length 20 kpc and radius 2 kpc.  
From the APEC spectral model, we find a best-fit electron
density of $n_e \approx 0.002$~cm$^{-3}$. We emphasize 
that the normalization is poorly constrained from these data (Figure 
\ref{fig:showfit}, but it is interesting to perform this estimate 
nevertheless. Taking $t \approx 10^{7}$ yr
and the range of mechanical luminosity from Paper I, we find $L_{\rm
  X} \approx 2 \times 10^{41} - 10^{42}$~erg~s$^{-1}$, in agreement
with what we observe.  The corresponding X-ray emitting gas mass is
$M_{\rm X} \approx 10^{10} f^{1/2}$~\msun, with a total thermal energy
of $E = PV = 2 n_e kT V \approx 1.3 \times 10^{58}\,f^{1/2}$ erg, or
$\dot{E} = 3 \times 10^{43}\, f^{1/2}$~erg~s$^{-1}$.  The
uncertainties in these calculations are quite large, due to the
uncertainties in the spectral fitting and 
the unknown volume and volume filling factor.  However, the kinetic
luminosity needed to power the X-ray outflow is within a factor of
three of what we inferred from the ionized gas outflow.

On the other hand, models of superwinds suggest that the wind is
powered by far more tenuous and hotter gas than we observe here, $\sim
10^7$~K for starbursts \citep{stricklandstevens2000} and perhaps even
hotter for AGN \citep[e.g.,][]{zubovasking2012}. In that case, the soft
X-rays may come from the surfaces of clouds as they are shocked by the
wind, and the X-ray luminosity cannot be inferred directly from the 
mechanical energy estimates. However, we can still phenomenologically 
compare the X-rays that we observe with other known superwinds.  

Most of the bolometric luminosity of starbursts and obscured quasars
is radiated at infrared wavelengths. Therefore, it is useful to compare
extended soft X-ray to total infrared flux ratios for a variety of
wind-driving systems. In starburst galaxies over a wide range in mass,
including ULIRGs, \citet{grimesetal2005} find $f_{\rm X}/f_{\rm FIR}
\approx 10 ^{-4}$. The ratio in \bub\ is consistent with this
value. Likewise the size scales as one might naively expect from the
infrared luminosity. The one way in which the X-ray gas observed here
differs significantly from that observed in starbursts is the
temperature.  Most of the ULIRGs in the Grimes et al.\ sample have
temperatures of \emph{kT}$\approx 600-800$ eV, as compared with the
\emph{kT}$\approx 280$ eV observed here.  Based on our fitting,
  we rule out a temperature of 600 eV at greater than 10~$\sigma$
  confidence.  This low inferred temperature may be a clue that we
are instead seeing photoionized gas.
 
\subsection{Photoionization}

Since we know that the central AGN is photoionizing gas on large
scales, and because of the correspondence in morphology between the
soft X-ray and warm ionized gas, we lastly consider the possibility
that what appears as X-ray continuum at our low spectral resolution is
actually composed of photoionized line emission.  Detailed analysis of
the extended X-ray emission around local Seyfert galaxies has found
strong evidence that the soft X-ray emission is dominated by
photoionization on large scales
\citep[][]{bianchietal2006,wangetal2011} and comprised predominantly
of line emission \citep[e.g.,][]{sambrunaetal2001}, although on galaxy
scales collisional ionization is also important
\citep[e.g.,][]{paggietal2012,wangetal2014}.  Based on these works,
the $L_{\rm X}/L_{\rm [OIII]}$ ranges from $0.1-0.3$, and is fairly
constant with radius. The relatively low temperatures of \emph{kT}$\approx
280$ eV are consistent with the temperatures that result from thermal
fits to other Seyfert galaxies. Also, the observed ratio of X-ray to
\oiii\ luminosity of $0.05-0.1$ is consistent with expectations from
photoionized gas (and is identical to that seen in NGC 4051 by Wang et
al. 2011). Furthermore, the orientation of the X-ray emission (N-S)
aligns with the direction of quasar illumination inferred from the
position angle in polarimetric observations (Paper I).

\section{Summary}

We have detected soft X-rays from an expanding bubble of warm ionized
gas in the obscured quasar and ULIRG \bub.  We discuss various origins
for the X-ray emission, concluding that the most probable are
photoionization and/or shocks from a quasar-driven superwind. To
determine the real origin of this emission requires deeper X-ray
observations.  As in \citet{bianchietal2006}, we hope to distinguish
between the spectrum of a thermally emitting gas, as expected in the
wind picture, from the line-dominated spectrum expected if
photoionization excites the gas.  In the wind picture, we expect that
there is a hotter and more tenuous wind component to search for,
although the expected temperature and luminosity is quite model
dependent.  In addition, the morphology of the soft X-rays will
provide further clues to its origin. In the hot wind scenario, the
X-rays will be uniformly distributed within the bubbles, whereas in
the photoionization scenario, the X-rays will trace the distribution
of \oiii\ emission in the bubbles.

\acknowledgements 
We thank the referee for a prompt and useful report. 
We thank C.~A. Faucher-Giguere, E. Quataert, and
J. P. Ostriker for inspiring us to look for X-ray emission from \bub.
J. M. C. acknowledges support from \chandra\ grant G02-13130.

\end{document}